\documentstyle[psfig]{mn}
\def\ts{\thinspace} \textheight 230mm
\topmargin -0.5cm \title[]{A Possible
Explanation for the Radio Afterglow of
GRB{980519}:  The Dense Medium Effect}

\author[]{X. Y. Wang$^{1}$, Z.G. Dai$^{1,2}$ and T. Lu$^{1,2}$ 
\thanks{E-mail: tlu@nju.edu.cn} \\
\rm $^1$Department of Astronomy, Nanjing University, Nanjing 210093, 
         P. R. China \\
\rm $^2$LCRHEA, Institute for High-Energy Physics, Chinese Academy of 
         Sciences, Beijing 100039, P. R. China} 
\date{Accepted ........
      Received .......;
      in original form .......}
\pubyear{2000}

\begin{document}

\maketitle 

\begin{abstract}
 
 GRB{980519} is characterized by its rapidly declining optical
 and X-ray afterglows. Explanations of this behavior include
 models invoking a dense medium environment which  makes the shock wave
 evolve quickly into the sub-relativistic phase, a jet-like outflow,
 and a wind-shaped circumburst medium environment. 
 Recently, Frail {et al}. (1999a) found that
 the latter two cases are consistent with the radio afterglow of this burst.
 Here, by considering the trans-relativistic shock hydrodynamics, we show
 that the dense medium model can also account for the radio light
 curve quite well. The potential virtue of the dense medium model
 for GRB{980519} is that it implies a smaller angular size of
 the afterglow, which is essential for interpreting the strong modulation
 of the radio light curve. 
  Optical extinction due to the dense medium is not important if the 
 prompt optical-UV flash accompanying the $\gamma$-ray emission can destroy
 dust by sublimation out to an appreciable distance.
 Comparisons with  some other radio afterglows
 are also discussed.
\end{abstract}
\begin{keywords}
gamma-rays: bursts  -- hydrodynamics: shock waves
\end{keywords}

\section{Introduction}
In the standard model of gamma-ray bursts (GRBs) (see Piran 1999 for
a review), an afterglow is generally believed to be produced by the
synchrotron radiation or inverse Compton scattering of the 
shock-accelerated electrons in an ultra-relativistic shock wave
expanding in a homogeneous medium. As more and more ambient matter
is swept up, the shock gradually decelerates while the emission from
such a shock fades down, dominating at the beginning in X-rays and
progressively at the optical to radio energy bands 
({M\'esz\'aros} \& {Rees} 1997; Waxman 1997a;
Wijers \& Galama 1999). 
In general, the light curves of X-ray and optical afterglows are
expected to exhibit power-law decays (i.e. $F_\nu\propto{t}^{-\alpha}$)
with the temporal index $\alpha$ in the range $1.1-1.4$, given
the energy spectral index of electrons $p\sim{2-3}$. The observations of the 
earliest afterglows are in good agreement with this simple model
(e.g. Wijers, Rees \& M\'esz\'aros 1997; 
Waxman 1997b). However, over the past year, 
we have come to recognize a class of GRBs whose afterglows showed
light curve breaks ( e.g. GRB\ts{990123}, GRB\ts{990510}; 
Kulkarni {\it et al.} 1999a; Harrison {\it et al.} 1999) or
steeper temporal decays (i.e. $F_\nu\propto{t}^{-2}$; e.g. 
GRB\ts{980519}, GRB\ts{980326}; Bloom {\it et al.} 1999). 
Explanations for these behaviors
include three scenarios: 1) a jet-like relativistic shock has undergone
the transition from the spherical-like phase to a sideways-expansion phase
(Rhoads 1999),
as suggested by some authors (e.g. Sari, Piran \& Halpern 1999; 
Kulkarni {\it et al.} 1999a; Harrison {\it et al.} 1999).
2) the shock wave propagates in  a wind-shaped circumburst environment 
with the number density $n\propto{r}^{-2}$
( Dai \& Lu 1998; {M\'esz\'aros}, {Rees} \& Wijers 1998;
Chevalier \& Li 1999; Chevalier \& Li 2000; Li \& Chevalier 1999 );
3) a dense medium environment ($ n\sim 10^{5}-10^{6}{\rm cm^{-3}}$) 
makes the shock
wave evolve into the sub-relativistic phase after a short relativistic one
(Dai \& Lu 1999a,b).
In the last model, since an afterglow from the shock at the sub-relativistic
stage decays more rapidly than at the relativistic one, we will expect
a light curve break or a long-term steeper decay, depending on the
time when it begins to enter into the sub-relativistic stage. This
scenario has reasonably interpreted the  break in the R-band
afterglow of GRB\ts{990123} (Dai \& Lu 1999a) and the steep decays of the
X-ray and optical afterglows of GRB\ts{980519} (Dai \& Lu 1999b).

Recently, Frail {\it et al.} (1999a) tried to test the first two models
(the jet and wind 
cases) by means of the radio afterglow behavior of GRB\ts{980519} and
found that the wind model described it rather well. Due to the strong
modulation of the light curve, however, they could not draw a decisive
conclusion for the jet case. In this paper, we will examine the 
possibility of describing the evolution of the radio afterglow in terms of 
the dense medium model. Since this scenario involves the transition phase 
of the shock wave from the relativistic stage to 
the sub-relativistic, we have considered
the trans-relativistic shock hydrodynamics in the numerical study.
We first present the  asymptotic
result of the fitting of the radio data in section 2, and then the numerical
result in section 3.
In section 4, we show that the optical extinction due to the dense circumburst 
medium is not important, since the prompt optical-UV radiation, caused by the reverse shock,
can destroy the dust by sublimation out to a substantial distance, as proposed by
Waxman \& Draine (1999). 
 Finally, we give our discussions and conclusions.
\section {Asymptotic behavior of the radio afterglow in the sub-relativistic
stage}

GRB\ts{980519} is the second brightest GRB in the BeppoSAX sample.
Its optical afterglow measured since $\sim8.5$ hours after the burst
exhibited  rapid fading, consistent with $t^{-2.05\pm0.04}$ in  $BVRI$
(Halpern {\it et al.} 1999; Djorgovski {\it et al.} 1998),
 and the power-law decay slope of the X-ray
afterglow, $\alpha_X=2.07\pm0.11$ (Owens {\it et al.} 1998),
 was in agreement with
the optical. The spectrum in optical band alone is well fitted by a power-law
$\nu^{-1.20\pm0.25}$, while the optical to X-ray spectrum can also be fitted by
a single power-law $\nu^{-1.05\pm0.10}$. The radio emission was observed
with the Very Large Array (VLA) (Frail {\it et al.} 1999a)
 since about 7.2 hours after 
the burst and referred as VLA J232221.5+771543. The radio light curve
shows  a gradual rise to  a plateau followed by a decline until below detectability
after about 60 days. There are some large
variations in these data, which is believed to be caused 
by interstellar scattering and 
scintillation (ISS; Frail {\it et al.} 1999a). 

As discussed by Dai \& Lu (1999b), the steep decays of the X-ray 
and optical afterglows
of GRB\ts{980519} can be attributed to the shock evolution into the
sub-relativistic phase 8 hours after the burst as the result of the 
dense circumburst medium. During such a sub-relativistic expansion phase, 
the hydrodynamics of the shocked shell is described by the self-similar
Sedov-von Neumann-Taylor solution. The shell radius and its velocity
scale with time as $r\propto{t_\oplus^{2/5}}$ and $\beta\propto{t_\oplus
^{-3/5}}$, where $t_\oplus$ denotes the time measured in the observer frame.
Then, we obtain the synchrotron peak frequency $\nu_m\propto{t_\oplus^{-3}}$,
the cooling frequency $\nu_c\propto{t_\oplus^{-1/5}}$, 
the peak flux $F_{\nu_m}\propto{t_\oplus^{3/5}}$ and the self-absorption
frequency $\nu_a\propto{t_\oplus^{-\frac{3p-2}{p+4}}}$ for the case of
$\nu_a>\nu_m$(Dai \& Lu 1999b). 
Now, the derived spectra and light curves are
\small{
\begin{equation}
F_\nu=\left \{
       \begin{array}{llllll}
        (\nu_a/\nu_m)^{-(p-1)/2}(\nu/\nu_a)^{5/2}F_{\nu_m}
         \propto \nu^{5/2}t_\oplus^{11/10},\\
         \hskip 3.6cm {\rm if}~~~~\nu_m<\nu<\nu_a ;\\
         (\nu/\nu_m)^{-(p-1)/2}F_{\nu_m} \propto \nu^{-(p-1)/2}
         t_\oplus^{(21-15p)/10}, \\
         \hskip 3.6cm {\rm if}~~~~\nu_a< \nu <\nu_c ;\\
         (\nu_c/\nu_m)^{-(p-1)/2}(\nu/\nu_c)^{-p/2}F_{\nu_m}\propto
              \nu^{-p/2}t_\oplus^{(4-3p)/2}, \\
         \hskip 3.6cm {\rm if}~~~~\nu>\nu_c.
        \end{array}
       \right.
\end{equation}
}
If the observed optical afterglow was emitted by
slow-cooling electrons while the X-ray afterglow from fast-cooling
electrons and if $p\approx 2.8$, then according to Eq.(1), 
the decay index $\alpha_R
=(21-15p)/10\approx -2.1$ and $\alpha_X=(4-3p)/2\approx -2.2$,
in excellent agreement with observations. Also, the model spectral
index at the optical to X-ray band, $\beta=-(p-1)/2 \approx
-0.9$ is quite consistent with the observed one
$-1.05\pm 0.10$. Furthermore, from the information
of X-ray and optical afterglows, Dai \& Lu (1999b) have inferred the 
physical parameters of this burst as follows:
\begin{equation}
\begin{array}{ll}
E\sim0.3\times10^{52}{\rm erg},~~~ \epsilon_e\sim0.16,~~~ \epsilon_B\sim2.8\times
10^{-4},\\
 n\sim3\times10^{5}{\rm cm^{-3}},~~~~~ z\sim0.55,
\end{array} 
\end{equation}
where $E$ is shock energy, $z$ is the redshift of the burst and $\epsilon_e$
and $\epsilon_B$ are the electron and magnetic energy 
fractions of the shocked medium, respectively. 

After the 60-day radio observational data being published, we promptly
checked the dense medium model, and found that the asymptotic analysis can
approximately describe the radio behavior. The analysis is as follows:
adopting the inferred values of the physical parameters in Eq.(2),
the detected frequency $\nu_{*}=8.46 {\rm GHz}$  equals to $\nu_a$ at about day 12;
thus, according to Eq.(1), we expect that before this time the radio emissions
rise as ${t_\oplus^{1.1}}$ and then decay as ${t_\oplus^{-2.1}}$ after
  the self-absorption frequency $\nu_a$ falls below $\nu_{*}$.
This simple asymptotic solution agrees qualitatively with  observations,
as showed in Fig. 1 by the dotted line
This preliminary analysis stimulated us to fit the radio data with
 a more detailed model by
taking into account the  trans-relativistic shock hydrodynamics and
the strict self-absorption effects of the synchrotron radiation.
\section{Trans-relativistic shock hydrodynamics, self-absorption effect
 and the fitting of the radio data }

We consider an instantaneous release of a large amount of energy $E$
in a constant density external medium. The energy released drives in the 
medium a shock wave, whose dynamic evolution from the relativistic
to sub-relativistic phase can be described approximately in the following way.

Let $r$ be the shock radius, $\gamma$ and $\Gamma$ be, respectively,
the Lorentz factors of the shell and the shock front, $\beta$ be the velocity
of the shock front. As usual, the shock expansion is assumed to be adiabatic,
during which the energy is conserved, and we have (Blandford \& McKee 1976)
\begin{equation}
\frac{4}{3}\pi\sigma{\beta}^{2}{\Gamma}^{2}r^{3}nm_{p}c^{2}=E,
\end{equation}
where $\sigma$ is a coefficient: $\sigma\rightarrow 0.35$ when $\beta\rightarrow
1$ and $\sigma\rightarrow 0.73$ when $\beta\rightarrow 0$.
As Huang, Dai \& Lu (1998), we use an approximate expression for $\sigma$:
$\sigma=0.73-0.38\beta$.

The radius of the shock wave evolves as (Huang, Dai \& Lu 1998)
\begin{equation}
\frac{dr}{dt_\oplus}=\beta{c}\gamma(\gamma+\sqrt{\gamma^2-1})/(1+z).
\end{equation}
and the jump conditions of the shock are given by (Blandford \& McKee 1976)
\begin{equation}
n'=\frac{\hat{\gamma}\gamma+1}{\gamma-1}n,~~~e'=\frac{\hat{\gamma}\gamma+1}
{\hat{\gamma}-1}(\gamma-1)nm_p{c^2},
\end{equation}
\begin{equation}
\Gamma^2=\frac{(\gamma+1)[\hat{\gamma}(\gamma-1)+1]^2}{\hat{\gamma}(2-
\hat{\gamma})(\gamma-1)+2},
\end{equation}
where $e'$ and $n'$ are the energy and the number densities of the shell in its
comoving frame and $\hat{\gamma}$ is the adiabatic
index, which equals $4/3$ for ultra-relativistic shocks and $5/3$ for
sub-relativistic shocks.
A simple interpolation between these two limits
$\hat{\gamma}=\frac{4\gamma+1}{3\gamma}$ 
gives a valid approximation
for trans-relativistic shocks (Dai, Huang \& Lu 1999).
 Using the above equations, we can now numerically obtain the evolution
 of $r(t_\oplus)$ and $\gamma(t_\oplus)$ 
 in the trans-relativistic stage, given proper initial conditions.

 As usual, we assume that the distribution of relativistic electrons
 with the Lorentz factor $\gamma_e$ takes a power-law form with the
 number density given by $n(\gamma_e)d\gamma_e=C{\gamma_e}^{-p}d\gamma_e$
 above a low limit $\gamma_{min}$, which is determined by the shock velocity:
 $\gamma_{min}=\epsilon_e\frac{(p-2)}{(p-1)}\frac{m_p}{m_e}(\gamma-1)$.
 Also, the energy densities of electrons and magnetic fields 
 are assumed to be proportional to the total energy density $e'$ in the comving
 frame as $U'_e=\epsilon_e{e'}$ and $B'_{\perp}=(8\pi\epsilon_B{e'})^{1/2}$.
 Thus, from the standard theory of the synchrotron radiation
  (Rybicki \& Lightman 1979; Li \& Chevalier 1999),
we have the expressions of the effective optical depth
 and the self-absorbed flux  
 \begin{equation}
 \tau_{\nu'}=\frac{p+2}{8\pi{m}{\nu'}^2}\frac{\sqrt{3}q^3}{2mc^2}(\frac
 {4\pi{mc}{\nu'}}{3q})^{-p/2}F_2(\frac{\nu'}{{\nu'_m}})C{B'_{\perp}}^{(p+2)/2}
 \Delta{r'},
\end{equation} 
\begin{equation}
\nu'_m=\frac{3{\gamma_{min}^{2}}qB'_{\perp}}{4\pi{mc}},~~~
C=(p-1)n'{\gamma_{min}^{p-1}},~~~ \Delta{r'}=r/{\eta},
\end{equation}
\begin{equation}
 F_{\nu}=(1+z)D^3\pi(\frac{r}{d_{L}})^2\frac{2m{\nu'}^2}{p+2}(\frac{4\pi{mc\nu'}}
 {3qB'_{\perp}})^{1/2}\frac{F_1(\nu'/\nu'_m)}{F_2(\nu'/\nu'_m)}
 (1-e^{-\tau}),
\end{equation}
where $F_1(x)$, $F_2(x)$ are defined by Eq.(5) in 
Li \& Chevalier (1999), $m$ and $q$
denote the mass and charge of the electron, $d_{L}$ is the luminosity distance
of the burst, assuming a flat Friedman universe with $H_0=65~ 
{\rm km~s}^{-1}{\rm ~Mpc}^{-1}$
 and $\eta\sim10$, characterizing the width of the shock shell.
Here $D\equiv{1/[\gamma(1-\beta)]}$ describes the relativistic effect,
and 
$\nu$ relates to the corresponding frequency $\nu'$ in the comoving frame 
by $\nu={D\nu'}/{(1+z)}$.

Using the above full set of equations, we computed the radio flux at the
 frequency $\nu_{*}=8.46{\rm GHz}$ and plotted the model fit in Fig. 1 as 
the solid line. We find that the following combination of the parameters
fits almost all valid data rather well: $E\sim0.8\times10^{52} {\rm erg}$, 
$\epsilon_e\sim0.2$, $\epsilon_B\sim 1\times10^{-4}$, 
$n\sim 1\times10^{5} {\rm cm^{-3}}$,
$z\sim0.55$ and the Lorentz factor at the initial time $t_\oplus=1/3~ {\rm days}$,
$\gamma=1.2$ ($\beta\sim0.55$). We stress that these parameters are in
excellent agreement with those inferred independently from the X-ray
and optical afterglows (Dai \& Lu 1999b), as listed in Eq.(2). 
Clearly, there are some large amplitude variations in the observed
light curve (e.g. about 18 days after the burst),
which is believed to be caused by diffractive scintillation.
Also plotted
in Fig. 1 is the fit (dashed line) computed with the sub-relativistic
model as presented in the appendix of Frail, Waxman \& Kulkarni (1999). 
The fit with this model
was obtained by adopting the initial conditions as: $t_0=1/3~ {\rm days}$
 and $r_0=1.6\times10^{16}{\rm cm}$ and the parameter values ($\epsilon_e$,
 $\epsilon_B$, $E$, $z$ and $n$ ) the same as those in Dai \& Lu (1999b).
Comparing the trans-relativistic model with sub-relativistic one, we can
easily see that the relativistic effect (as characterized by 
$D\equiv{1}/{\gamma(1-\beta)}$ in Eq.(9)) flattens the rising phase at
 earlier time, making the trans-relativistic model agree
better with the observations, while at the later
 time both the fitting curves trend towards the asymptotic solution (i.e.
$F_\nu\propto{t_\oplus^{-2.1}}$).
 
\begin{figure}
\centerline{\hbox{\psfig{figure=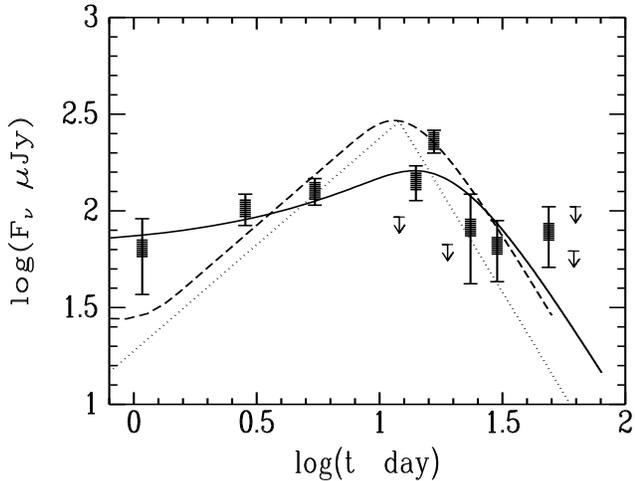,width=4.5in,angle=270}}}
\caption {Model fits of the radio light curve at 8.46 GHz.  
Detections and upper limits for the non-detections,
 taken from Frail {\it et al.} (1999a),
 are indicated by the filled squares and arrows, respectively.
The dotted line represents the expected asymptotic solution of the radio
behavior according to the dense medium model  in 
Dai \& Lu (1999b). The solid and dashed lines
represent the fits with the tran-relativistic and sub-relativistic models,
respectively.
See text for additional details.}
\end{figure}

\section{Dust sublimation and optical extinction by dense medium}

One may ask whether the dense circumburst medium  may cause large extinction in the 
optical afterglow of GRB\ts{980519}. A crude estimate is as follows.
At the time that the blast wave transits to the sub-relativistic stage 
($t_\oplus\sim\frac13{\rm days}$, $\beta\sim0.55$, $\gamma\sim1.2$),
from Eq.(3) we derived its radius to be $r\sim2.1\times10^{16}{\rm cm}$.
Therefore the characteristic column density through the medium into which
the blast wave is expanding is about $nr\sim2.1\times10^{21}{\rm cm^{-2}}$ with
the corresponding $A_{V}$ of 1.3 magnitudes in the rest frame of the absorber.
This column density is comparable to (but slightly larger than) the Galactic 21cm
column density ($\sim1.74\times10^{21}{\rm cm^{-2}}$, Halpern et al. 1999) in the 
direction of the GRB\ts{980519}. With $A_{\lambda}$ scales linearly with $\frac{1}
{\lambda}$, the absorption in our observed B band for this optical transient
is about $1.3(1+z)\sim2$ magnitudes, twice the value adopted by Halpern et al. (1999)
for correction of the relative extinction.

In the above estimate, we have made a questionable assumption, i.e. the
dense medium around the burst has a standard gas-to-dust ratio. However, this
may be not realistic, considering that the dust around the burst can be destroyed due to 
sublimation out to an appreciable distance ($\sim$ a few pc) by the
prompt optical-UV flash (Waxman \& Draine 1999;
hereafter WG99), accompanying the prompt burst.
Below we will give an estimation of the destruction radius for dense medium case,
following WG99.

The prompt optical flash detected accompanying GRB\ts{990123} (Akerlof et al. 1999) suggests
that, at least for some GRBs, $\gamma$-ray emission is accompanied by prompt
optical-UV radiation with luminosity in the 1-7.5eV range of the order of 
$10^{49}(\frac{\Delta\Omega}{4\pi})\rm erg/s$ for typical GRB parameters, where
$\Delta\Omega$ is the solid angle into which $\gamma$-ray and optical-UV emission is 
beamed (WG99). The most natural explanation of this flash is emission from a reverse shock
propagating into the fireball ejecta shortly after it interacts with the surrounding 
gas (Sari \& Piran 1999; {M\'esz\'aros} \& {Rees} 1999). As for GRB\ts{980519} with the
parameter values as $\epsilon_e\sim0.2$,  $\epsilon_B\sim10^{-4}$,
$n\sim10^{5}\rm cm^{-3}$, $E\sim8\times10^{51}\rm erg$ and the burst duration
$\Delta{t}\sim70 \rm {s}$, we derive the luminosity in the 1-7.5 eV range to be
about $L_{1-7.5}\sim5\times10^{48}\rm erg/s$ (Here we have assumed that electron
and magnetic field energy fractions in the reverse shock are similar to those
in the forward shock; Wang, Dai \& Lu 1999c). The condition for the grain to be
completely sublimed during the prompt flash time is 
\begin{equation}
T>{T_c}\simeq2300{\rm K}[1+0.033{\rm ln}(\frac{a_{-5}}{\Delta{t}/10{\rm s}})],
\end{equation}
where $T$ is the grain temperature, determined by Eq.(8) of WG99, and $a\equiv{a_{-5}
\times{10^{-5}\rm cm}}$ is the 
radius of the dust grain. Then, according to Eq.(17) of WG99, the radius out to which the
prompt flash can heat grains to the critical temperature $T_c$ is 
\begin{equation}
R_c\simeq3.7\times10^{19}(\frac{Q_{UV}L_{49}(1+0.1a_{-5})}{a_{-5}})^{1/2}{\rm cm}
\simeq2.7\times10^{19} {\rm cm},
\end{equation}
where $Q_{UV}$ is the absorption efficiency factor of the optical-UV flash and
can be assumed to be near one for grain radii $a>10^{-5}$ expected in dense medium.
Thus, we can safely say that the extinction due to the circumburst dense medium is
not important if the size of the dense medium is shorter than $R_c$ and we think that
this condition is reasonable for GRB\ts{980519}, and  also for GRB\ts{990123}
(Dai \& Lu 1999a).

\section{Discussions and Conclusions}

The detection of strong diffractive scintillation requires that the 
angular size of the source VLA\thinspace{J232221.5+771543} should
be less than $1\mu{\rm arcsec}$ 
even 15 days after the burst (Frail {\it et al.} 1999a), otherwise the fluctuations
would be suppressed. This small inferred size is not consistent with
the spherical, homogeneous model with a normal 
density ($n\sim1{\rm cm^{-3}}$), but
 can marginally be consistent with  the jet model and the wind-shaped circumburst
 medium model. We note that the dense medium model may have a potential 
advantage for this requirement, simply because that the shock will be
quickly decelerated to a sub-relativistic velocity and therefore
have a shorter shock radius. This can be clearly seen
from the following comparison: $\theta_{s,rel}\propto
\int_{0}^{t_\oplus}2\gamma^2{cdt}({1}/{\gamma(t_\oplus)})$ for the 
relativistic case while 
$\theta_{s,sub}\propto\int_{0}^{t_\oplus}\beta{c}{dt} (\beta<1)$ 
for the sub-relativistic one.
Here, for the afterglow of GRB\ts{980519}, we assume that before $t_\oplus
\sim1/3 {\rm days}$, the shock is adiabatic and relativistic; thus
the shock radius is $r(t)\simeq(17Et_{\oplus}/4\pi{m_p}nc(1+z))^{1/4}\simeq
0.78\times10^{16} (E/3\times10^{51}{\rm erg})^{1/4}
(n/3\times10^{5}{\rm cm^{-3}})^{-1/4}
(t_\oplus/{\frac{1}{3} \rm days})^{1/4}[(1+z)/1.55]^{-1/4} \rm cm$
(Sari, Piran \& Narayan 1998). 
Then, the shock radius should follow the Sedov-von Neumann-Taylor self-similar
solution as $r(t_\oplus)\propto{t_\oplus^{2/5}}$. Thus, we obtained the angular size
of the afterglow  $\theta_{s}\simeq
0.8 \mu{\rm arcsec}(t_\oplus/15 {\rm days})^{2/5}$. 
The agreement would be even improved if we note that
at the beginning of the assumed sub-relativistic stage, the radius 
should increase more slowly than the self-similar solution in the 
trans-relativistic regime.

The strong modulations caused by scintillation also make  the
estimate of the spectral slope in the radio band less accurate. 
The {\em averaging} value of the spectral slope from day 12 on 
is $\beta\simeq{-0.45\pm0.6}$ (where
$F_\nu\propto{\nu^{\beta}}$), implying that the time averaged self-absorption 
frequency
$\nu_a$ is between 1.43 GHz and 4.86 GHz (Frai {\it et al.} 1999b).
 In our model,
the time ($t_\oplus\simeq 12{\rm days}$) 
when the fitting curve begins to decline corresponds to 
$\nu_a=\nu_*=8.46\rm GHz$. Since the self-absorption
frequency decays as $\nu_a\propto{t_\oplus}^{-(3p-2)/(p+4)}
\propto{t_\oplus^{-0.94}}$ for $p=2.8$, we expect that $\nu_a$ shifted quickly 
below 4.86 GHz at day 21, but was above 1.43 GHz over all the detecting
time, which is in reasonable agreement with the observations.

The radio afterglow of GRB\ts{970508}, the longest light curve ($450$-day)
obtained by far, exhibited different behavior from GRB\ts{980519}
(Frail et al. 1997; Waxman, Kulkarni \& Frail 1998; 
Frail, Waxman \& Kulkarni 1999).
From the spectral and temporal radio behavoir, Frail, Waxman \& Kulkarni (1999)
inferred that the fireball has undergone a transition to sub-relativistic
expansion at $t\sim 100~{\rm days}$, consistent with the inferred low
ambient density $n\sim 1{\rm cm^{-3}}$ (but also see Chevalier \& Li 1999a). 
On the other hand, some radio afterglows (e.g. GRB\ts{990510}, 
GRB\ts{981226}; Frail {\it et al.} 1999b)
show similar behaviors to GRB\ts{980519}, that is, they exhibit a slow
rise to the maximum for a relatively short time 
and then a fast decline until below detectability. 
It is likely that the shocks of these bursts entered into the sub-relativistic
stage after a short relativistic one and our above model can also describe their
radio afterglows. Harrison {\it et al.} (1999) had interpreted the 
broad-band lightcurve
break in the afterglows of GRB\ts{990510} as due to a jet-like outflow.
We speculate that another possible 
explanation is that the shock had entered into the sub-relativistic stage
after $\sim 1{\rm days}$ as the result of the 
combination of the dense medium and jet effects (Wang {\it et al.} 1999b),
the latter of which may be real in consideration of the large inferred
isotropic energy.
The radio afterglow of GRB\ts{990123} is unique for its ``flare" behavior
(Kulkarni {\it et al.} 1999b), whose most natural explanation is that it arises from
the reverse shock, as evidenced by the prompt optical 
flash (Sari \& Piran 1999).
Our preliminary computation (using the trans-relativistic model) shows that
the radio emission from the forward shock in the dense medium model is
significantly lower than that from the reverse shock and declines quickly after
the peak time, if a jet-like outflow with an opening angle $\theta\sim0.2$,
as required by the ``energy crisis" of this burst, is 
invoked (Wang {\it et al.} 1999b).
Moreover, the fast decline of the radio emission from the forward shock,
which is caused by the deceleration of the shock in the sub-relativistic
stage, can be consistent with the non-detection even 3 days after the burst.

In summary, we argue that the dense medium model, which has interpreted
the optical to X-ray afterglows of GRB\ts{980519} quite well, can also
account for the radio afterglow excellently. The circumburst environment
can affect the evolution of GRBs afterglows significantly
 ({M\'esz\'aros}, {Rees} \& {Wijers} 1998;
{Panaitescu}, {Meszaros} \& {Rees} 1998; Wang {\it et al.} 1999a). For a low
($n\sim 1{\rm cm^{-3}}$), homogeneous density environment, the shock waves
stay at the relativistic shock stage for quite a long time, while for
the dense medium case, the shock wave
 quickly enters into the sub-relativistic stage.
Recently, a generic dynamic model for
the evolution of shocks from ultra-relativistic phase to the
 sub-relativistic one has been also developed by Huang {\it et al.} (1999a).
The afterglows of the optically thin radiation (e.g. optical and X-rays)
from the shock at the sub-relativistic stage decays more rapidly than
at the relativistic one. As for the radio afterglow 
(usually $\nu_{m}<{\nu_a}$
at the sub-relativistic stage for this model), the dense medium model 
predicts a slow rise ($\nu_*<\nu_a$), followed by a round peak and a
late steep decline ($\nu_*>\nu_a$), trending towards the behavior of 
the optical and X-ray afterglows. Clearly, this behavior is different
from the jet model in the early epoch. But it is somewhat similar to the wind
model, making it difficult to distinguish 
between them through the radio observations.

\section*{Acknowledgments}
 We would like to thank the referee Dr. R. Wijers for his valuable suggestions and improvment
on this manuscript. X. Y. Wang also thank
Dr. Y.F. Huang for helpful discussions.
This work was supported by the National Natural Science Foundation
of China under grants 19773007 and 19825109 and the Foundation 
of the Ministry of Education of China..


\begin{thebibliography}{}
\bibitem[]{}
 Akerlof, C. W. {\it et al.} 1999, Nature, 398, 400.

\bibitem[{Blandford} \& {McKee} 1976]{bm76}
{Blandford}, R. D. and {Mckee}, C. F. 1976, Phys. Fluids, 19, 1130.

\bibitem[Bloom {\it et al.} 1999]{blo99}
Bloom, J. S. {\it et al.}, 1999, Nature, 401, 453.

\bibitem[{Chevalier} \& {Li} 1999a]{cl99a}
{Chevalier}, R.~A. and {Li}, Z.-Y. 1999, ApJL, 520, L29.

\bibitem[{Chevalier} \& {Li} 1999b]{cl99b}
{Chevalier}, R.~A. and {Li}, Z.-Y. 2000, ApJ,  536, 195.

\bibitem [{Dai} \& Lu 1998]{dl98} 
{Dai}, Z. G. and {Lu}, T., 1998, MNRAS, 298, 87 

\bibitem [{Dai} \& Lu 1999a]{dl99a}
{Dai}, Z. G. and {Lu}, T., 1999a, ApJ, 519, L155.

\bibitem [{Dai} \& Lu 1999b]{dl99b}
{Dai}, Z. G. and {Lu}, T., 1999b, ApJ,  537,in press, astro-ph/9906109.

\bibitem [{Dai}, {Huang} \& {Lu} 1999]{dhl99}
{Dai}, Z. G., {Huang}, Y. F. and {Lu}, T. 1999, ApJ, 520, 634.

\bibitem[Djorgovski {\it et al.}  1998]{djo98}
Djorgovski, S.~G., Gal, R.~R., Kulkarni, S.~R., Bloom, J.~S., and Kelly, A.
  1998, GCN  79.
  
\bibitem[Frail {\it et al.} 1997] {fra97}
Frail, D. A., Kulkarni, S. R., Nicastro, L., Feroci, M., Taylor, G. B.
1997, Nature, 389, 261.  

\bibitem[Frail {\it et al.} 1999a] {fra99a}
Frail, D. A. {\it et al.}, 1999a, ApJ,  in press, astro-ph/9910060. 


\bibitem[Frail {\it et al.} 1999b] {fra99b}
Frail, D. A. {\it et al.}, 1999b, ApJL, 525,L81. 


\bibitem[Frail, Waxman \& Kulkarni 1999]{fwk99}
Frail, D. A., Waxman, E. and Kulkarni, S. R. 1999, ApJ, in press, 
astro-ph/9910319.

\bibitem[{Halpern} {\it et al.}  1999]{hal99}
{Halpern}, J.~P., {Kemp}, J., {Piran}, T., and {Bershady}, M.~A. 1999, ApJ,
  517, L105.

\bibitem[{Harrison} {\it et al.} 1999]{har99}
{Harrison, F. A.} {\it et al.}, 1999, ApJ, 523, L121.

\bibitem[{Kulkarni} {\it et al.} 1999a]{kul99a}
{Kulkarni, S. R.} {\it et al.}, 1999a, Nature, 398, 389.

\bibitem[{Kulkarni} {\it et al.} 1999b]{kul99b}
{Kulkarni, S. R.} {\it et al.}, 1999b, ApJ, 522, L97.

\bibitem[Huang, Dai \& Lu 1998]{hdl98}
Huang, Y. F., Dai, Z. G. and Lu, T. 1998, A\&A, 336, L69.


\bibitem[Huang, Dai \& Lu 1999]{hdl99a}
Huang, Y. F., Dai, Z. G. and Lu, T. 1999a, MNRAS, 309, 513.


\bibitem[Li \& Chevalier 1999]{lc99}
Li, Z.-Y. and Chevalier, R.~A. 1999,  ApJ, 526, 716.

\bibitem[{M\'esz\'aros}, {Rees} \& {Wijers} 1998]{mrw98}
{M\'esz\'aros}, P., {Rees}, M.~J., and {Wijers}, R. A. M.~J. 1998, ApJ, 499,
  301.

\bibitem[{M\'esz\'aros} \& {Rees} 1997]{mr97}
{M\'esz\'aros}, P. and {Rees}, M.~J. 1997, 476, 232.

\bibitem[]{}
 {\rm M$\acute{e}$sz$\acute{a}$ros} P.,  Rees M. J.,
1999, MNRAS, 306, L39.
  
  
\bibitem[Owens {\it et al.} 1998]{owe98}
Owens, A. {\it et al.} 1998, A\&A, 339, L37.  
  
\bibitem [{Piran} 1999]{pir99}
{Piran}, T., 1999, Phys. Rep., 314, 575.

\bibitem[{Panaitescu}, {Meszaros} \& {Rees} 1998]{pmr98}
{Panaitescu}, A., {Meszaros}, P., and {Rees}, M.~J. 1998, ApJ, 503, 314.

\bibitem[Rhoads 1999]{rho99}
Rhoads, J.~E. 1999, ApJ, 525, 737.

\bibitem[Rybicki \& Lightman 1979]{rl79}
Rybicki, G. B. and Lightman, A. P. 1979, Radiative Process in Astrophysics
(New York: Wiley).


\bibitem[{Sari}, {Piran} \& {Halpern} 1999]{sph99}
{Sari}, R., {Piran}, T., and {Halpern}, J.~P. 1999, ApJ, 519, L17.

\bibitem[{Sari}, {Piran} \& {Narayan} 1998]{spn98}
{Sari}, R., {Piran}, T., and {Narayan}, R. 1998, ApJ, 497, L17.

\bibitem[{Sari} \& {Piran} 1999]{sp99}
{Sari}, R., and {Piran}, T., 1999, ApJ, 517, L109.

\bibitem[Wang {\it et al.} 1999a]{wdl99a}
Wang, X. Y., Dai, Z. G., Lu, T., Wei, D. M. and Huang, Y. F., 1999a, A\&A, accepted,
astro-ph/9910029.

\bibitem[Wang {\it et al.} 1999b]{wdl99b}
Wang, X. Y., Dai, Z. G. and Lu, T., 1999b, in preparation.

\bibitem[]{}
 Wang, X. Y., Dai, Z. G. and Lu, T., 1999c,  MNRAS, accepted, astro-ph/9906062

\bibitem[Waxman 1997a]{wax97a}
Waxman, E. 1997a, ApJ, 489, L33.

\bibitem[Waxman 1997b]{wax97b}
Waxman, E. 1997b, ApJ, 491, L19.

\bibitem[Waxman, Kulkarni \& Frail 1998]{wkf98}
Waxman, E., Kulkarni, S. R. and Frail, D. A. 1998, ApJ, 497, 288.

\bibitem[]{}
 Waxman, E. and Draine, B. T., 1999, ApJ, in press, astro-ph/9909020.

\bibitem[Wijers, Rees \& M\'esz\'aros 1997]{wrm97}
Wijers, R. A. M.~J., Rees, M.~J., and M\'esz\'aros, P. 1997, MNRAS, 288, L51.

\bibitem[Wijers \& Galama 1999]{wg99}
Wijers, R. A. M.~J. and Galama, T. J. 1999, ApJ,
523, 177.

\end{thebibliography}
\end{document}